\documentclass{ifacconf}
\usepackage{graphicx}      
\usepackage{natbib}        
\usepackage{amsmath,amssymb,amsfonts}
\usepackage{tikz}
\usepackage[percent]{overpic}
\usetikzlibrary{positioning, arrows.meta}
\usepackage{xcolor}
\usepackage{algorithm}
\usepackage{algpseudocode}
\usepackage{url}

\begin{document}
\begin{frontmatter}

\title{A Nonlinear Observer for Air-Velocity and Attitude Estimation Using Pitot and Barometric Measurements\thanksref{footnoteinfo}} 

\thanks[footnoteinfo]{This work was supported by the"Grands Fonds Marins" Project Deep-C, and the ASTRID ANR project ASCAR. This research work is also supported in part by NSERC-DG RGPIN-2020-04759 and Fonds de recherche du Québec (FRQ).}

\author[First]{Nyoba Tchonkeu, Melone} 
\author[Second]{Berkane, Soulaimane} 
\author[Third]{Hamel, Tarek}

\address[First]{University of Quebec in Outaouais, Gatineau, QC J8X3X7, Canada, (e-mail: nyom01@uqo.ca).}
\address[Second]{University of Quebec in Outaouais, Gatineau, QC J8X3X7 and Lakehead University, Thunder Bay, ON P7B 5E1, Canada, (e-mail: soulaimane.berkane@uqo.ca).}
\address[Third]{I3S-UniCA-CNRS, University Cote d’Azur and the Insitut Universitaire de France,
        06903 Sophia Antipolis, France, (e-mail: thamel@i3s.unice.fr).}

\begin{abstract}
This paper addresses the problem of estimating air velocity and full attitude for unmanned aerial vehicles (UAVs) in GNSS-denied environments using minimal onboard sensing—an interesting and practically relevant challenge for UAV navigation. The contribution of the paper is twofold:
(i) an observability analysis establishing the conditions for uniform observability (UO), which are useful for trajectory planning and motion control of the UAV; and
(ii) the design of a nonlinear observer on $\mathrm{SO}(3)\ltimes\mathbb{R}^3 \times \mathbb{R}$ that incorporates pitot-tube, barometric altitude, and magnetometer measurements as outputs, with IMU data used as inputs, within a unified framework. Simulation results are presented to confirm the convergence and robustness of the proposed design, including under minimally excited trajectories.
\end{abstract}


\end{frontmatter}

\section{Introduction} \label{sec:intro}
Robust estimation of air velocity and attitude is fundamental for the safe and autonomous operation of UAVs. The challenge is particularly acute in GNSS-denied environments such as urban canyons or contested airspaces, where satellite-based signals are unavailable or unreliable. Under such constraints, UAVs must rely on onboard (typically lightweight) sensors, and the design of observers capable of delivering accurate estimates with minimal instrumentation remains a central research problem.

Several authors have addressed the attitude estimation problem from inertial measurement units (IMUs). Nonlinear complementary filter on \(\mathrm{SO}(3)\) \citep{mahony2008nonlinear}, geometric observers formulations \citep{lageman2008observer,barczyk2012invariant}, and symmetry-preserving observers approaches \citep{2008_bonnabel_SymmetryPreservingObservers} have demonstrated robust performance when aided by magnetometer or GNSS measurements. However, for fixed-wing UAVs, the accelerometer measures both gravity and aerodynamic forces, invalidating the gravity-only assumption and causing attitude observers to misinterpret dynamic loads as reference vectors \citep{mahony2008nonlinear}, thereby reducing robustness during maneuvering flight. This limitation has motivated the development of velocity-aided attitude (VAA) observers, which uses velocity information to enhance observability. Examples include observers using GNSS-derived inertial velocity \citep{2016_hua_StabilityAnalysisVelocityaided, hansen2017nonlinear,hua2017riccati, grip2013nonlinear,benallegue2023velocity,berkane2017attitude} or Doppler radar \citep{2013_troni_PreliminaryExperimentalEvaluation}. While effective, such methods are constrained by cost, hardware demands, and GNSS vulnerabilities, limiting their use on lightweight aerial platforms.

These constraints have increased the use of Pitot-based sensing in small UAVs. Pitot systems provide lightweight and direct measurements of specific components of the body-frame air velocity and are therefore standard equipment on fixed-wing platforms. Early approaches fused Pitot, IMU, and aerodynamic models \citep{lie2014synthetic, borup2016nonlinear}, but accuracy depended strongly on aerodynamic parameters that are difficult to calibrate for small airframes. To mitigate this sensitivity, recent works have incorporated Pitot measurements directly into observer design \citep{sun2019observability,johansen2015estimation}, thereby reducing reliance on uncertain aerodynamic models. Oliveira et al.~\citep{oliveira2024pitot} further advanced this direction by proposing a Riccati--nonlinear observer on $\mathrm{SO}(3)\times\mathbb{R}^3$ for joint air-velocity and the gravitational direction estimation in GNSS-denied flight, and establishing the first rigorous local stability guarantees for such architectures. Owing to the sensing configuration, the system retains a structural yaw ambiguity and satisfies uniform observability only under sufficiently rich persistent excitation.

On the other hand, barometer measurements are widely available on most UAV platforms and, when properly incorporated, can relax the observability requirements of Pitot–IMU–based architectures. To this end, we extend the nonlinear observer framework of \citet{oliveira2024pitot} by incorporating barometric altitude and magnetometer measurements. The barometer, routinely embedded in autopilot systems, provides continuous pressure-based altitude measurements and has long been used for vertical state estimation in fixed-wing UAVs \citep{mahony2011fixedwing}. It is commonly fused with GNSS/INS for height stabilization \citep{borup2016nonlinear}, and recent studies have indicated its potential to enhance attitude estimation in GNSS-denied conditions \citep{tchonkeu2025barometer}. The magnetometer supplies a global directional reference, enabling full attitude estimation \citep{Bryne2017,tchonkeu2025barometer}. In particular, we extend the system with a scalar altitude state that couples the original dynamics (attitude and air velocity) with the barometric measurement. We assume that the vertical wind velocity is negligible, which is standard in the absence of dedicated wind–sensing instrumentation and allows the barometric measurement to be consistently incorporated into the model. We then propose a nonlinear observer, based on the Riccati–observer framework of \citet{hamel2017riccati}, whose linearized error dynamics admit a bounded state matrix. Under suitable observability conditions, this property enables establishing local exponential stability of the estimation errors.

The remainder of the paper is organized as follows. Section~\ref{sec:prelims} introduces notation and mathematical preliminaries. Section~\ref{sec:prob_desc} formulates the estimation problem. Section~\ref{sec:obs_design} presents the proposed observer, while Section~\ref{sec:obs_stab_analysis} provides the associated observability and stability analysis. Implementation aspects are discussed in Section~\ref{sec:implementation}, followed by simulation results in Section~\ref{sec:simulation} and concluding remarks in Section~\ref{sec:concl}.

\section{Preliminary Material} \label{sec:prelims}
We denote by \( \mathbb{R} \) and \( \mathbb{R}_+ \) the sets of real and nonnegative real numbers, respectively. The \( n \)-dimensional Euclidean space is denoted by \( \mathbb{R}^n \). The Euclidean inner product of two vectors \( x, y \in \mathbb{R}^n \) is defined as \( \langle x, y \rangle = x^\top y \). The associated Euclidean norm of a vector \( x \in \mathbb{R}^n \) is \( |x| = \sqrt{x^\top x} \). Furthermore, we denote by \( \mathbb{R}^{m \times n} \) the set of real \( m \times n \) matrices. The set of \( n \times n \) positive definite matrices is denoted by \( \mathcal{S}^+(n) \), and the identity matrix is denoted by \( I_n \in \mathbb{R}^{n \times n} \). Given two matrices \( A, B \in \mathbb{R}^{m \times n} \), the Euclidean matrix inner product is defined as \( \langle A, B \rangle = \mathrm{tr}(A^\top B) \), and the Frobenius norm of \( A \in \mathbb{R}^{n \times n} \) is given by \( \|A\| = \sqrt{\langle A, A \rangle} \). 
The unit sphere \( \mathbb{S}^2 := \{ \eta \in \mathbb{R}^3 \mid |\eta| = 1 \} \subset \mathbb{R}^3 \) denotes the set of unit 3D vectors and forms a smooth submanifold of \( \mathbb{R}^3 \).
The special orthogonal group of 3D rotations is denoted by
$
\mathrm{SO}(3) := \{ R \in \mathbb{R}^{3 \times 3} \mid RR^\top = R^\top R = I_3,\ \det(R) = 1 \}.
$
The Lie algebra of $\mathrm{SO}(3)$ is 
$
\mathfrak{so}(3) := \{\, \Omega \in \mathbb{R}^{3\times 3} \mid \Omega^\top = -\Omega \,\},
$
isomorphic to $\mathbb{R}^3$ via the skew-symmetric operator 
$(\cdot)^\times : \mathbb{R}^3 \to \mathfrak{so}(3)$, defined such that 
$
u \times v = u^\times v$ for all $u,v \in \mathbb{R}^3.$
The exponential map \( \exp : \mathfrak{so}(3) \rightarrow \mathrm{SO}(3) \) defines a local diffeomorphism from a neighborhood of \( 0 \in \mathfrak{so}(3) \) to a neighborhood of \( I_3 \in \mathrm{SO}(3) \). This enables the composition map \( \exp \circ (\cdot)^\times : \mathbb{R}^3 \rightarrow \mathrm{SO}(3) \), which is given by the following Rodrigues' formula \citep{Ma2004rodriguesformula}:
\begin{equation}
\exp([\theta]^{^\times}) 
= I_3 - \frac{\sin(\|\theta\|)}{\|\theta\|}[\theta]^{^\times}
+ \frac{1-\cos(\|\theta\|)}{\|\theta\|^2}([\theta]^{^\times})^2.
\label{eq:rodriguesformula}
\end{equation}
\section{Problem Description}\label{sec:prob_desc}
\begin{figure}[ht]
  \centering
  \begin{overpic}[width=0.7\linewidth]{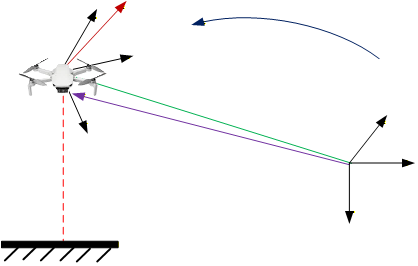}
    \put(11,44){\small \(\left\{\mathcal{B}\right\}\)}
    \put(22,62){\small \(e_1^{\mathcal{B}}\)}
    \put(30,52){\small \(e_2^{\mathcal{B}}\)}
    \put(20,28){\small \(e_3^{\mathcal{B}}\)}
    \put(18,20){\small \(h\)}
    \put(78,28){\small \(\left\{\mathcal{I}\right\}\)}
    \put(92,37){\small \(e_1\)}
    \put(95,27){\small \(e_2\)}
    \put(82,7){\small \(e_3\)}
    \put(50,28){\small \(p\)}
    \put(66,52){\small \(R^\top\)}
    \put(31,62){\small \(V_a\)}
    \put(40,40){\small \(v\)}
  \end{overpic}
  \caption{We assume a vehicle equipped with an IMU, a set of $m$ Pitot tubes measuring components of the air velocity $V_a$, and a barometer measuring the height $h$.
  }
  \label{fig:veh_baro_imu}
\end{figure}
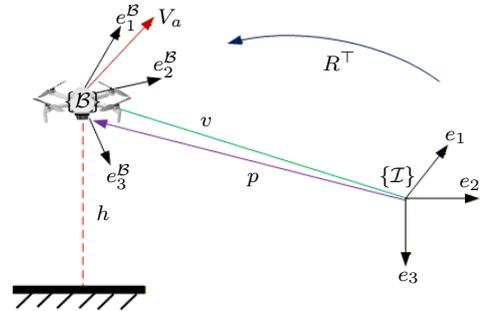
Consider a fixed-wing aerial vehicle equipped with an IMU (gyroscopes and accelerometers), a set of $m$ Pitot probes, and a barometric pressure sensor (see fig.~\ref{fig:veh_baro_imu}).

Let \( R \in \mathrm{SO}(3) \) denote the rotation matrix from the vehicle body-fixed frame \( \mathcal{B} \) to the inertial frame \( \mathcal{I} \), and let \( \boldsymbol{\omega} \in \mathbb{R}^3 \) and  \( \mathbf{a} \in \mathbb{R}^3 \) be the body angular velocity and the linear specific acceleration measured by IMU, respectively. The position and linear velocity of the rigid body, expressed in the inertial frame \( \mathcal{I} \), are denoted by \( p \in \mathbb{R}^3 \) and \( v \in \mathbb{R}^3 \), respectively. The gravitational acceleration is expressed in the inertial frame by \( \mathbf{g} = g \mathbf{e}_3 \in \mathbb{R}^3 \), where \( \mathbf{e}_3 = \begin{bmatrix}0  & 0 & 1 \end{bmatrix}^\top \in \mathbb{S}^2 \) denotes the standard gravity direction, and \( g \approx 9.81\, \text{m/s}^2 \) is the gravity constant. The vehicle's translational and rotational kinematics are given by
\begin{align*}
\dot{v} &= R\mathbf{a} + g \mathbf{e}_3, \\
\dot{R} &= R\boldsymbol{\omega}^\times.
\end{align*}
Let \( v_w \in \mathbb{R}^3 \) be the inertial wind velocity. The vehicle air-velocity in inertial and body frames are \(v_a = v - v_w \) and \(V_a = R^{\top}v_a\), respectively. Assuming a constant bounded wind velocity throughout the flight ( \(\dot v_w \approx 0 \)), one obtains \(\dot v_a = \dot v\). Hence, the air-velocity and attitude dynamics become :
\begin{align}
\dot{v}_a &= R\mathbf{a} + g \mathbf{e}_3, \label{eq:air_velocity_dyn} \\
\dot{R} &= R\boldsymbol{\omega}^\times.
\label{eq:attitude_dyn}
\end{align}
The vehicle is equipped with $m$ calibrated Pitot probes. Each probe measures the scalar projection of the air-velocity $V_a$ along a known body-fixed direction \(b_i \in \mathbb{S}^2\):
\begin{equation} 
y_{p,i} = b_i^\top V_a + n_{p,i} = b_i^\top R^Tv_a + n_{p,i},
\end{equation}
where $n_{p,i} \sim \mathcal{N}(0,\sigma_{p,i}^2)\in \mathbb{R}$ is zero-mean noise. Collectively, multi-probe Pitot system vector is modeled as
\begin{equation}
y_p = B^\top V_a + n_p, \quad B = [b_1\ \cdots b_m] \in \mathbb{R}^{3\times m}, \label{eq:pitot_measurement}
\end{equation}
where \(n_p = [n_{p,1}\ \cdots n_{p,m}]^\top \in \mathbb{R}^m\).
To enhance the observability of the vehicle's motion, a scalar altitude variable is introduced as 
$h := \mathbf{e}_3^\top p \in \mathbb{R}$, which is measured by the barometric sensor modeled:
\begin{equation}
y_b = h + n_b,
\label{eq:baro_measurement}
\end{equation}
where $n_b \sim \mathcal{N}(0,\sigma_b^2) \in \mathbb{R}$ is zero-mean noise. 

We also assume that the vertical wind velocity, $e_3^\top v_w$, is negligible, as it is typically much smaller than horizontal winds in low-altitude UAV operations.
Furthermore, we assume that the IMU includes a magnetometer. The measured Earth's magnetic field in the body-frame \(\mathbf{m}_{\mathcal{B}}  \in \mathbb{S}^2\) is then modeled as
\begin{equation}
\mathbf{m}_{\mathcal{B}} = R^\top m_{\mathcal{I}}+n_m,
\label{eq:mag_measurement}
\end{equation}
where $m_{\mathcal{I}} \in \mathbb{S}^2$ is the known magnetic field vector expressed in the inertial frame and $n_m \sim \mathcal{N}(0,\sigma_m^2) \in \mathbb{R}^3$ is zero-mean noise. 
The objective is to estimate the full air-velocity and attitude states from these measurements, exploiting the altitude and heading information to enhance the overall system observability and estimation performance of the coupled air-velocity--attitude dynamics described by~\eqref{eq:air_velocity_dyn} and~\eqref{eq:attitude_dyn}.

\section{Proposed Observer Design}\label{sec:obs_design}

\begin{figure}[ht]
  \centering
  \begin{overpic}[width=0.9\linewidth]{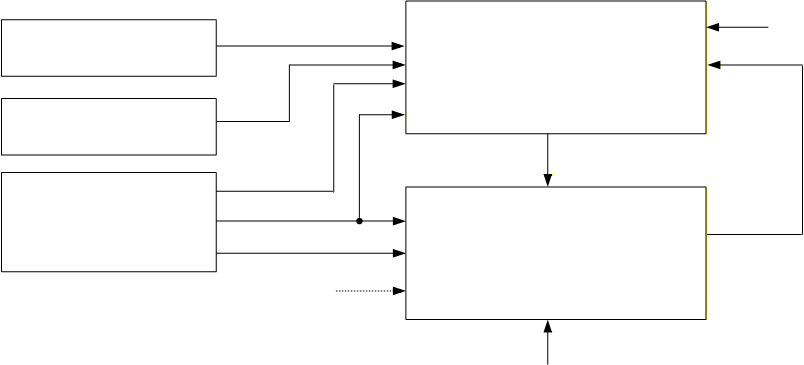}
    \put(2,39){\small Pitot tubes}
    \put(2,29){\small Barometer}
    \put(8,18){\small IMU }
    \put(32,24){\small \(\mathbf{m}_{\mathcal{B}}\)}
    \put(40,19){\small \(\mathbf{a}\)}
    \put(40,15){\small \(\boldsymbol{\omega}\)}
    \put(32,32){\small \(h\)}
    \put(32,41){\small \(V_a\)}
    \put(28,8){\small \(v_w (\approx 0)\)}
    \put(51,40){\small Riccati Equation~\eqref{eq:cre}}
    \put(51,34){\small Innovation terms~\eqref{eq:innovation_inputs}}
    \put(96,42){\small \(m_{\mathcal{I}}\)}
    \put(60,24){\small \(\delta_R,\)}
    \put(72,24){\small \(\delta_v,\ \delta_h\)}
    \put(58,16){\small Nonlinear}
    \put(58,11){\small Observer}
    \put(62,7){\small~\eqref{eq:observer_dyn}}
    \put(90,18){\small\(\hat{R}\)}
    \put(88,11){\small \(\hat{V}_a, \hat{h}\)}
    \put(65,0){\small \( \mathbf{g}\)}
  \end{overpic}
\caption{Overall structure of the proposed nonlinear observer.}
  \label{fig:observer_block_diagram}
\end{figure}
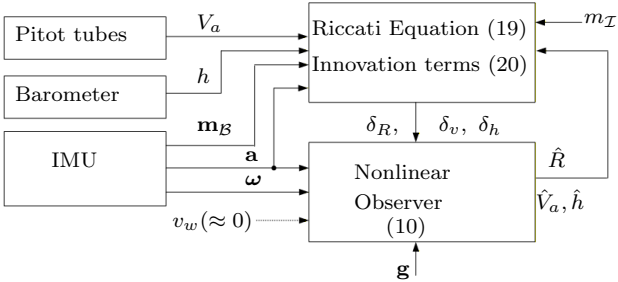

This section presents the design of a barometric-augmented observer that extends the local framework of \citet{oliveira2024pitot} to include barometric and magnetic measurements.
The proposed architecture comprises a local Riccati equation~\eqref{eq:cre} that processes the linearized attitude--air-velocity--altitude error dynamics to generate innovation terms~\eqref{eq:innovation_inputs}, which are then injected into a nonlinear observer~\eqref{eq:observer_dyn} on \( \mathrm{SO}(3) \ltimes \mathbb{R}^3 \times \mathbb{R} \) to estimate the full orientation, air-velocity, and altitude. The overall architecture is illustrated in fig.~\ref{fig:observer_block_diagram}.

From~\eqref{eq:air_velocity_dyn} and~\eqref{eq:attitude_dyn}, the body-frame air-velocity dynamics follow directly from \citet{oliveira2024pitot} and are given by:
\begin{equation}
\dot{V}_a = -\boldsymbol{\omega}^\times V_a +gR^\top \mathbf{e}_3 + \mathbf{a}.
\label{eq:Va_dot}
\end{equation}
The inertial altitude is defined as $h := \mathbf{e}_3^\top p$. Since $\dot p = v \approx v_a$, it follows that
\begin{equation}
\dot h = \mathbf{e}_3^\top v_a = \mathbf{e}_3^\top R V_a.
\label{eq:h_dot}
\end{equation}
Combining~\eqref{eq:attitude_dyn} and \eqref{eq:Va_dot}--\eqref{eq:h_dot}, we propose the following nonlinear observer on $\mathrm{SO}(3) \ltimes \mathbb{R}^3 \times \mathbb{R}$, whose structure consists of a copy of the system dynamics augmented with suitable innovation terms:
\begin{equation}
\begin{cases}
\dot{\hat R} =\hat R\boldsymbol{\omega}^\times - \delta_R^\times\hat R,\quad \hat{R}(0) \in \mathrm{SO}(3),\\
\dot{\hat V}_a = -\boldsymbol{\omega}^\times \hat V_a +g\hat R^\top \mathbf{e}_3 + \mathbf{a} -\hat R^{\top}\delta_v + \hat R^{\top}\delta_R^\times\hat R\hat V_a  ,\\
\dot{\hat h} = \mathbf{e}_3^\top \hat R \hat V_a -\delta_h, \label{eq:observer_dyn}
\end{cases}
\end{equation}
where $\hat R \in \mathrm{SO}(3)$ denotes the estimate of \(R\), \(\hat V_a \in \mathbb{R}^3\) the estimate of \(V_a\), and \(\hat h \in \mathbb{R}\) the estimate of \(h\).
\(\delta_R \in \mathbb{R}^3,~\delta_v \in \mathbb{R}^3,~\delta_h \in \mathbb{R}\) are the correction terms, which will be determined from the linearization process below.

Let \(\tilde R:= R\hat R^\top\), \(\tilde v_a:= RV_a - \hat{R}\hat V_a\), and \(\tilde h:= h - \hat h\) denote, respectively, the attitude, inertial air-velocity, and barometric errors. From~\eqref{eq:attitude_dyn} and \eqref{eq:Va_dot}--\eqref{eq:observer_dyn}, one obtains the following nonlinear error dynamics:
\begin{subequations} \label{eq:error_dyn}
\begin{align}
\dot{\tilde R} &= \tilde R\delta_R^\times \label{eq:R_error_dyn}, \\
\dot{\tilde v}_a &= (\tilde{R}-I_3)\hat R\mathbf{a}+\delta_v \label{eq:va_error_dyn}, \\
\dot{\tilde h} &= \mathbf{e}_3^{\top}\tilde{v}_a + \delta_h, \label{eq:h_error_dyn}
\end{align}
\end{subequations}
To linearize~\eqref{eq:error_dyn}, we express \(\tilde{R}\) in terms of its associated unit quaternion \(\boldsymbol{\tilde q} :=(\tilde q_0,~\tilde q) \), using Rodrigues' formula \citep{Ma2004rodriguesformula}:
\begin{equation}
\tilde{R} = I_3 + 2\tilde{q}^\times(\tilde q_0I_3+\tilde{q}^\times), \label{eq:R_err_quat}
\end{equation}
where for a rotation by an angle $\theta$ about a unit axis $\mathrm{\boldsymbol{u}} \in \mathbb{S}^2$, one has $\tilde q_0 = \cos({\theta/2})$ and $\tilde q = \mathrm{\boldsymbol{u}}\sin({\theta/2})$. For small attitude errors ($\delta\theta$), $\|\tilde{q}\|\ll 1$ and $\tilde q = \mathrm{\boldsymbol{u}}{\delta\theta/2}$. Then, one obtains the second-order approximation \(\tilde{q}_0 \approx 1 + \mathcal{O}(\|\tilde{q}\|^2)\). Substituting this into~\eqref{eq:R_err_quat}, yields
\[
\tilde{R} = I_3 + (2\text{sign}(\tilde{q}_0)\tilde{q})^\times + \mathcal{O}(\|\tilde{q}\|^2). \]
By defining $\tilde{\lambda} = 2\text{sign}(\tilde{q}_0)\tilde{q} \in \mathbb{R}^3$ with \(\|\tilde{\lambda}\| = \mathcal{O}(\|\tilde{q}\|)\), the first order approximation of attitude error becomes
\begin{equation}
\tilde{R} = I_3 + \tilde{\lambda}^\times +\mathcal{O}(\|\tilde{\lambda}\|^2)\label{eq:R_err_quat_first_order}.
\end{equation}
Exploiting the above first order approximation, one can rewrite \eqref{eq:error_dyn}, as follows:
\begin{subequations} \label{eq:error_dyn_first_order}
\begin{align}
\dot{\tilde{\lambda}} &= \delta_R + \mathcal{O}(\|\tilde{\lambda}\|\|\delta_R\|),\label{eq:err_small_angle_dyn} \\
\dot{\tilde v}_a &= -(\hat{R}\mathbf{a})^{\times}\tilde{\lambda} +\delta_v + \mathcal{O}(\|\tilde{\lambda}\|^2) \label{eq:dot_va_err}, \\
\dot{\tilde h}&=\mathbf{e}_3^{\top}\tilde{v}_a + \delta_h \label{eq:dot_h_err}.
\end{align}
\end{subequations}
For the output, we define the estimated Pitot, barometer, and magnetometer measurements, respectively, as 
$\hat y_p := B^\top \hat V_a$, $\hat y_b := \hat h$, and $\hat y_m := \hat R \mathbf{m}_{\mathcal{B}}$. 
In view of~\eqref{eq:pitot_measurement}--\eqref{eq:mag_measurement} and 
\eqref{eq:R_err_quat_first_order}, the output errors 
$\tilde y_p := y_p - \hat y_p$, 
$\tilde y_b := y_b - \hat y_b$, and 
$\tilde y_m := m_{\mathcal{I}} - \hat y_m$
satisfy
\begin{align*}
\tilde y_p &= B^{\top}\hat R^{\top}(\hat R\hat V_a)^{\times}\tilde\lambda
             + B^{\top}\hat R^{\top}\tilde v_a
             + \mathcal{O}(\|\tilde\lambda\|\|\tilde v_a\| + \|\tilde\lambda\|^2), \\
\tilde y_m &= -(m_{\mathcal{I}})^{\times}\tilde\lambda
             + \mathcal{O}(\|\tilde\lambda\|^2),\\
\tilde y_b &= \tilde h.
\end{align*}

Let 
$x := [\,\tilde\lambda^{\top}\;\tilde v_a^{\top}\;\tilde h\,]^{\top} \in\mathbb{R}^7$, 
$u(t) := [\,\delta_R^{\top}\;\delta_v^{\top}\;\delta_h\,]^{\top}$, and 
$y := [\,\tilde y_p^{\top}\;\tilde y_m^{\top}\;\tilde y_b\,]^{\top}\in\mathbb{R}^{m+4}$.
The local error dynamics admit the linear time-varying (LTV) representation
\begin{subequations}\label{eq:ltv}
\begin{align}
\dot x &= A(t)x + u(t)
         + \mathcal{O}(\|x_1\|\|u\| + \|x_1\|^2), \label{eq:ltv_dot_x}\\
y &= C(t)x
     + \mathcal{O}(\|x_1\|\|x_2\| + \|x_1\|^2). \label{eq:ltv_C_est}
\end{align}
\end{subequations}

The matrices $A(t)$ and $C(t)$ in~\eqref{eq:ltv} are given explicitly by
\begin{equation}
A(\hat R(t)) =
\begin{bmatrix}
\mathbf{0}_{3\times 3} & \mathbf{0}_{3\times 3} & \mathbf{0}_{3\times 1} \\
-(\hat R\mathbf{a})^{\times} & \mathbf{0}_{3\times 3} & \mathbf{0}_{3\times 1} \\
\mathbf{0}_{1\times 3} & \mathbf{e}_3^{\top} & 0
\end{bmatrix},
\label{eq:state_matrix}
\end{equation}

\begin{equation}
C(\hat R(t), \hat V_a(t)) =
\begin{bmatrix}
B^{\top}\hat R^{\top}(\hat R\hat V_a)^{\times} & B^{\top}\hat R^{\top} & \mathbf{0}_{m\times 1} \\
-(m_{\mathcal{I}})^{\times} & \mathbf{0}_{3\times 3} & \mathbf{0}_{3\times 1} \\
\mathbf{0}_{1\times 3} & \mathbf{0}_{1\times 3} & 1
\end{bmatrix}.
\label{eq:est_output_matrix}
\end{equation}

To determine the innovation terms $\delta_{R}$, $\delta_v$, and $\delta_h$, 
we design a local Kalman--Bucy-type gain $K(t)$ for~\eqref{eq:ltv}. 
Using the innovation input $u(t) = -K(t)y(t)$, the closed-loop error dynamics become
\begin{multline}
\dot{x}
= (A - K(t)C(t))x
+ \mathcal{O}(\|x_1\|\|Ky\| + \|x_1\|^2)\\
  - K(t)\,\mathcal{O}(\|x_1\|\|x_2\| + \|x_1\|^2),
\label{eq:riccati_observer}
\end{multline}
where $K(t) = P(t)C^\top(t)Q \in \mathbb{R}^{7\times (m+4)}$ and 
$P(t)$ solves the continuous-time Riccati equation (CRE)
\begin{equation}
\dot{P}
= A(t)P + PA^\top(t) - PC^\top(t)Q\,C(t)P + S,
\label{eq:cre}
\end{equation}
with $P(0)>0$. The matrices $Q>0$ and $S>0$ are symmetric positive definite, 
where $Q$ encodes the sensor precisions ($Q_p$ for the Pitot, $Q_m$ for the magnetometer, 
and $Q_b$ for the barometer), while $S$ plays the role of the process noise covariance.
We decompose
\[
K(t) =
\begin{bmatrix}
K_{\delta_R}(t)^{\top} &
K_{\delta_v}(t)^{\top} &
K_{\delta_h}(t)^{\top}
\end{bmatrix}^{\top},
\]
with 
$K_{\delta_R}\in \mathbb{R}^{3\times(m+4)}$, 
$K_{\delta_v}\in \mathbb{R}^{3\times(m+4)}$, and 
$K_{\delta_h}\in \mathbb{R}^{1\times(m+4)}$. 
Then, using the definition of $u(t)$, 
the Riccati solution yields the innovation terms
\begin{subequations}\label{eq:innovation_inputs}
\begin{align}
\delta_R(t) &= -\,K_{\delta_R}(t)\,y(t), \\
\delta_v(t) &= -\,K_{\delta_v}(t)\,y(t), \\
\delta_h(t) &= -\,K_{\delta_h}(t)\,y(t).
\end{align}
\end{subequations}

The stability and convergence properties of~\eqref{eq:innovation_inputs} are fundamentally linked to the well-posedness of the CRE, which requires the UO of the associated LTV model. Let us define the matrices $A^{\star}(t) = A(R(t))$ and $C^{\star}(t)= C(R(t), V_a(t))$ associated with the true trajectory. The UO of the linearized error system in~\eqref{eq:ltv} inherites from the UO of $(A^\star (t), C^\star (t))$. According to Theorem.~3.1 of \cite{hamel2017riccati}, if the pair $(A^{\star}(t),C^{\star}(t))$ is uniformly observable, the CRE in~\eqref{eq:cre} admits a unique, bounded, and positive-definite solution \( P(t) \) for all \( t \ge 0 \). As a consequence, the local error $x$ of ~\eqref{eq:ltv} decays exponentially to the origin, with the rate of convergence tuned by \(Q\) larger than some positive matrix and \(S>0\), ensuring stability and convergence of the nonlinear observer in~\eqref{eq:observer_dyn}.

\section{Observability and Stability Analysis}\label{sec:obs_stab_analysis}
Consider the LTV system in~\eqref{eq:ltv}. By definition from \citet{Besancon2007}, the system ~\eqref{eq:ltv} or pair \( (A(t), C(t)) \) is \emph{uniformly observable} if there exist constants 
\( \delta, \mu > 0 \) such that, for all \( t \geq 0 \),
\begin{equation}
W(t, t+\delta) := \frac{1}{\delta} \int_t^{t+\delta} 
\Phi^\top(s, t) \, C^\top(s) \, C(s) \, \Phi(s, t) \, ds 
\geq \mu I_n, \label{eq:observability_gramian}
\end{equation}
where \( \Phi(s, t) \) is the state transition matrix such that
\begin{equation}
\frac{d}{dt} \Phi(s, t) = A(t)\Phi(s, t), 
\quad \Phi(t, t) = I_n, \quad \forall s \geq t. \label{eq:transition_matrix} 
\end{equation}
\( W(t, t+\delta) \) is called the \emph{observability Gramian} of the system.

The purpose of the following lemma is to establish trajectory-dependent sufficient conditions for the uniform observability of the pair $(A^\star, C^\star)$ associated with the true system. Hence, for a given trajectory, these conditions can be verified without explicitly computing the corresponding observability Gramian.
\begin{lem}\label{Lemma1}
Assume that the body-frame linear acceleration $\mathbf{a}(t)$, the angular velocity $\boldsymbol{\omega}(t)$, and the air-velocity $V_a(t)$ are continuous and uniformly bounded. Moreover, assume that vectors $m_\mathcal{I}$ and $\mathbf{e}_3$ are non-collinear. Let $J=[\mathbf{e}_1\ \mathbf{e}_2]\in\mathbb{R}^{3\times2}$, where $\mathbf{e}_1$ and $\mathbf{e}_2$ denote the first two canonical basis vectors of $\mathbb{R}^3$. Define $\Pi(t)=B^\top R(t)J$ and $a_\pi(t)=\mathbf{e}_3^\top(R(t)\mathbf{a}(t))^\times J$.
Assume there exist $\bar\delta>0$ and $\bar\mu_1,\bar\mu_2>0$ such that, for all $t\ge0$,
\begin{equation}
\frac{1}{\bar\delta}\int_t^{t+\bar\delta}\Pi(s)^\top\Pi(s)\,ds
\ \ge\ \bar\mu_1 I_2,
\label{eq:PE_Pi}
\end{equation}
and
\begin{equation}
\frac{1}{\bar\delta}\int_t^{t+\bar\delta}a_\pi(s)^\top a_\pi(s)\,ds
\ \ge\ \bar\mu_2 I_2.
\label{eq:PE_a_pi}
\end{equation}
Then the time-varying pair $(A^\star(t),C^\star(t))$ is uniformly observable.
\end{lem}
\begin{pf}
See Appendix~\ref{appendix_B}.
\end{pf}
The excitation requirements of Lemma~\ref{Lemma1} differ fundamentally from the Pitot-only setting of \cite{oliveira2024pitot}.  
With a single Pitot probe, \cite{oliveira2024pitot} requires both yaw and pitch excitation to ensure observability, since the PE condition is imposed on the full matrix  $B^\top R^\top$ and no absolute heading is available.  
In contrast, the present formulation—with magnetometer and barometer—separates the excitation across sensing channels.  
The Pitot contribution enters only through the horizontal projection $B^\top R(t)J$, so with $b_1 = \mathbf{e}_1$ the required motion reduces to \emph{yaw excitation only}.  
Moreover, when two probes are available with linearly independent horizontal projections, $B^\top R(t)J$ already has full rank in the horizontal plane, and the Pitot-induced PE condition can be satisfied \emph{without any motion}, whereas the Pitot-only case of \cite{oliveira2024pitot} requires three probes for the same effect.  
Finally, the baro-induced term $a_\pi(t)$ contributes vertical excitation whenever the inertial acceleration possesses a horizontal component—purely vertical acceleration does not help.  
Altogether, the PE conditions \eqref{eq:PE_Pi}--\eqref{eq:PE_a_pi} reorganize the needed excitation across horizontal Pitot geometry, horizontal components of the inertial specific force, and absolute yaw from the magnetometer.  
The following theorem follows directly from this uniform observability property.
\begin{thm}\label{Theorem1}
Consider the system~\eqref{eq:air_velocity_dyn}-\eqref{eq:mag_measurement} along with the closed loop error~\eqref{eq:riccati_observer} and the nonlinear observer~\eqref{eq:observer_dyn}, with the innovation inputs~\eqref{eq:innovation_inputs}, and $P(t)$ the symmetric positive definite matrix solution to the CRE~\eqref{eq:cre}. Assume the pair \((A^\star(t), C^\star(t))\) is uniformly observable, then the origin of the closed error dynamic~\eqref{eq:riccati_observer} is locally exponentially stable.
\end{thm}

\begin{pf}
Consider the Lyapunov function \(L:=x^\top P^{-1}x\). Using the CRE in~\eqref{eq:cre} and the innovation terms in~\eqref{eq:innovation_inputs}, one obtains (see \cite{sastry1999lyapunov}) 
\begin{equation}
\begin{aligned}
\dot L =& -\tfrac12\,x^\top\!\big(C^\top(t)Q\,C(t)+P^{-1}(t)S\,P^{-1}(t)\big)x\\
&+ \mathcal{O}(\|x\|^3,\|P\|^{-2}).
\label{eq:dotL}    
\end{aligned}
\end{equation}
The quadratic term in~\eqref{eq:dotL} is strictly dissipative provided \(P(t)\) and \(P^{-1}(t)\) are uniformly bounded and positive definite. By Lemma~\ref{Lemma1}, the pair \((A(t), C(t))\) in~\eqref{eq:ltv} is uniformly observable. Hence, along the equilibrium trajectory (i.e., when \(\Pi(t) = B^{\top} R(t) J\, \text{and}\, a_{\pi}(t) = \mathbf{e}_3^{\top} (R(t)\mathbf{a}(t))^{\times}J\)), the associated Riccati solution is bounded and well-conditioned (see Prop.~17 of \cite{oliveira2024pitot}). Furthermore, by Theorem 3.1 of Hamel and Samson, 2017 or Prop.~25 of \cite{barrau2019linear}, it follows that for \(Q >0\) and \(S >0\), the solution \(P(t)\) of the CRE in~\eqref{eq:cre} exists along bounded trajectories and remains uniformly bounded and positive definite. As a result, both \(P(t)\) and \(P^{-1}(t)\) are uniformly bounded, so the remainder term \(\mathcal{O}(\|x\|^3,\|P\|^{-2} )\) can be dominated by the strictly negative quadratic term in a sufficiently small neighborhood of the origin. Consequently, there exists \(\gamma >0\) such that \(\dot L \le -\gamma L\), which implies
\(
\|x(t)\| \le ce^{-\gamma(t-t_0)}\|x(t_0)\|, \quad \forall t\ge t_0,
\) for some \(c>0\).
Hence, the origin is a locally exponentially stable equilibrium for the closed-loop error dynamics in~\eqref{eq:riccati_observer}.
\end{pf}

Theorem~\ref{Theorem1} ensures local exponential convergence whenever the excitation conditions of Lemma~\ref{Lemma1} are satisfied.  
When $m\ge 3$, the probe geometry provides sufficient horizontal excitation through $B^\top R(t)J$ without requiring specific vehicle motion (see \cite{oliveira2024pitot}).  
For $m=2$, convergence can also occur without motion when the two probes have linearly independent horizontal projections, since the vertical direction is recovered through the baro–acceleration coupling.  
When only a single probe is available, horizontal excitation must arise from the vehicle motion, and in this case, yaw excitation is sufficient thanks to the magnetometer and barometric measurements.  
Thus, exponential convergence may be achieved either through probe geometry or through horizontal excitation induced by the vehicle dynamics.

\section{Discrete Time Implementation}\label{sec:implementation}
The proposed observer is implemented at the IMU sampling period~$T$.  Over each interval $[t_k,t_{k+1}]$, we assume the measured acceleration $\mathbf a_k$, the angular velocity $\boldsymbol\omega_k$, and the body-frame air-velocity $V_{a,k}$ are constant 
(see \cite{Bryne2017}). Moreover, we assume the discrete process noise $S_{d,k}\approx S_k T$, the estimated air-velocity \(\hat V_a (t) \approx\hat V_{a,k}\), and the estimated attitude \(\hat R(t) \approx \hat R_k\) for \( t \in [t_k,t_{k+1}]\).
Using these assumptions, a first-order discretization of the continuous-time state matrix $A(t)$  yields
\begin{equation}
A_{d,k} \approx
\begin{bmatrix}
I_3 & \mathbf{0}_{3 \times 3}  & \mathbf{0}_{3\times 1} \\
-T(\hat R_k \mathbf{a}_k)^\times &  I_3 & \mathbf{0}_{3\times 1} \\
\mathbf{0}_{1\times 3}& T\mathbf{e}_{3}^{\top} &1
\end{bmatrix}. \label{eq:Ad}
\end{equation}
The continuous-time output matrix $C(t)$ is sampled at the measurement timestamp as follows
\begin{equation}
C_{k} = \begin{bmatrix}C_{p,k}^\top,C_{m,k}^\top, C_{b,k}^\top\end{bmatrix}^\top \label{eq:Ck}
\end{equation}
where
\begin{equation}
 C_{p,k}
= \begin{bmatrix}
B^{\top}\hat{R}_k^{\top}(\hat{R}_k\hat{V}_{a,k})^{\times} & B^\top \hat R_k^\top  & \mathbf{0}_{m \times 1}
\end{bmatrix}, \label{eq:cpk}
\end{equation}
\begin{equation}
 C_{m,k}
= \begin{bmatrix}
-(m_{\mathcal{I}})^{\times}& \mathbf{0}_{3 \times 3} & \mathbf{0}_{3 \times 1}
\end{bmatrix}. \label{eq:cmk}
\end{equation}
\begin{equation}
 C_{b,k}
= \begin{bmatrix}
\mathbf{0}_{1 \times 3} & \mathbf{0}_{1 \times 3} &1
\end{bmatrix}. \label{eq:cbk}
\end{equation}
The resulting discrete-time observer, summarized in Algorithm~\ref{algo:Discrete_proposed_obs}, 
follows the standard correction--prediction structure with the above state and output matrices. The nonlinear observer~\eqref{eq:observer_dyn} is discretized at the 
IMU frequency using exponential-Euler and Euler integrations on $\mathrm{SO}(3)\times\mathbb{R}^3 \times \mathbb{R}$ \citep{mahony2008nonlinear}(see lines~28--32 of Algorithm~\ref{algo:Discrete_proposed_obs}).
\begin{algorithm}[!t]
\caption{Discrete-Time Implementation of the proposed observer}
\label{algo:Discrete_proposed_obs}
\begin{algorithmic}[1]
\Statex \textbf{Inputs:} $x_{0|0}, P_{0|0},\hat R_0, \hat V_{a,0}, \hat h_{0}$; $g_k,\mathbf{a}_k,\boldsymbol{\omega}_k, \mathbf{m}_{\mathcal{B}_k}$; $y_{p,k}$; $y_{m,k}$; $y_{b,k}$; $T$; $Q_k$; $S_k$; $v_w \approx 0$; $B = b_1$.
\Statex \textbf{Outputs:} $\,\hat R_{k},\,\hat V_{a,k}, \, \hat h_k, \,$ \(\forall k \in \mathbb{N}\)
\For{each time \(k \geq 0\)}
\If{IMU data \(\mathbf a_k\), \(\omega_k\) is available}
    \State {$A_{d,k} \gets $~\eqref{eq:Ad};} \quad $S_{d,k}\gets S_k\,T$;
    \State{\(P_{k+1|k} \gets A_{d,k} P_{k|k} A_{d,k}^\top + S_{d,k}.\)}
\EndIf
\If{$y_{p,k}$, $y_{m,k}$, and $y_{b,k}$ are all available}
\State $C_{p,k} \gets$~\eqref{eq:cpk}; $C_{m,k}~\gets$ \eqref{eq:cmk}; $C_{b,k}~\gets$ \eqref{eq:cbk}; $C_k \gets$~\eqref{eq:Ck};
\State $y_k\gets \begin{bmatrix} y_{p,k} - B^\top \hat V_{a,k}\\ y_{m,k} - \hat R_k \mathbf{m}_{\mathcal{B}_k} \\ y_{b,k} - \hat h_{k}
\end{bmatrix};$ 
\State $Q_k \gets \mathrm{blkdiag}(Q_{p,k},Q_{m,k},Q_{b,k})$.
\ElsIf{Barometer measurement is available}
\State $C_{b,k} \gets$~\eqref{eq:cbk}; $C_k \gets C_{b,k}$;
\State \( y_{k} \gets y_{b,k} - \hat h_{k}\); $Q_{k} \gets Q_{b,k}$.
\ElsIf{Magnetometer measurement is available}
\State $C_{m,k} \gets$~\eqref{eq:cmk}; $C_k \gets C_{m,k}$;
\State \( y_{k} \gets  y_{m,k} - \hat R_k \mathbf{m}_{\mathcal{B}_k}\); $Q_{k} \gets Q_{m,k}$.
\ElsIf{Pitot measurement is available}
\State $C_{p,k} \gets$~\eqref{eq:cpk}; $C_k \gets C_{p,k}$;
\State \(y_{k} \gets y_{p,k} - B^\top \hat V_{a,k}\); $Q_{k} \gets Q_{p,k}$.
\EndIf
\If{$y_{p,k}$ or $y_{b,k}$ or $y_{m,k}$ or all are available}
\State $K_k = P_{k+1|k}\,C_k^\top \big(C_k P_{k+1|k} C_k^\top + Q_k\big)^{-1}$;
\State $P_{k+1|k+1}=\left(I_7 - K_k C_k \right)P_{k+1|k}$;
\State $u_k \gets-K_k y_k$; $\delta_{R_{k}} \gets u_k(1{:}3)$; $\delta_{v,k} \gets u_k(4{:}6)$; $\delta_{h,k} \gets u_k(7)$.
\Else
    \State $u_k\gets 0$;$P_{k+1|k+1}\gets P_{k+1|k}$;$\delta_{R,k},\delta_{v,k},\delta_{h,k} \gets 0$. 
\EndIf
\State $P_{k+1|k+1}\gets \frac{1}{2}(P_{k+1|k+1} + P_{k+1|lso k+1}^{\top})$
\State $\hat R_{k+1} \gets \hat R_k \exp\!\big((\boldsymbol{\omega}_k - \hat R_k^\top \delta_{R,k})^\times T\big)$;
\State $\zeta_{1,k} \gets -\boldsymbol{\omega}_k^\times \hat V_{a,k} +g\hat R_k^\top \mathbf{e}_3 + \mathbf{a}_k$
\State $\zeta_{2,k} \gets -\hat R_k^{\top}\delta_{v,k} + \hat R_k^{\top}\delta_{R,k}^\times\hat R_k\hat V_{a,k}$
\State $\hat V_{a, k+1} \gets \hat V_{a,k}+T\left(\zeta_{1,k}+\zeta_{2,k} \right)$;
\State $\hat h_{k+1} \gets \hat h_k + T\left(\mathbf{e}^{\top}_3\hat R_k\hat V_{a,k} - \delta_{h,k} \right)$
\EndFor
\end{algorithmic}
\end{algorithm}

\section{Simulation Results}\label{sec:simulation}
To evaluate the performance of the proposed observer, we conduct a simulation of a vehicle illustrated in fig.~\ref{fig:veh_baro_imu} equipped with one calibrated Pitot probe and the barometer-IMU system moving in a 3D space.
The ground truth inertial velocity and altitude are respectively given by \(v(t) = [-1.5\sin(1.5t),3\cos(3t),-15\sqrt{3}\cos(3t)/4]^\top\) and \(h(t) = -5\sqrt{3}\sin(3t)/4\).
We excite only the yaw component. Hence, the angular velocity is given by
\(
\boldsymbol{\omega}(t) =[ 0, 0,0.7 \sin(1.6 t)]^\top
\)
and the body-frame linear acceleration is given by
\[
\mathbf{a}(t) =R^\top\begin{bmatrix}- 2.25\cos(1.5t)\\
- 9\sin(3t)\\
11.25\sqrt{3}\sin(3t) -g  
\end{bmatrix}.
\]
The pitot geometry is such that \(B= b_1 = \mathbf{e}_1\). We assume the absence of wind \(v_w \approx 0\) in that 3D space. This implies that the ground truth inertial air-velocity is given by \(v_a \approx v\). Hence, the ground truth body-frame air-velocity is generated by \(V_a = R^\top v\). The constant vector $m_\mathcal{I}$ is set to $m_\mathcal{I} =\begin{bmatrix} 1/\sqrt{2} & 0& 1/\sqrt{2} \end{bmatrix}^{\top}$. 

\begin{figure}[!t]
    \centering
    \includegraphics[width=0.8\linewidth]{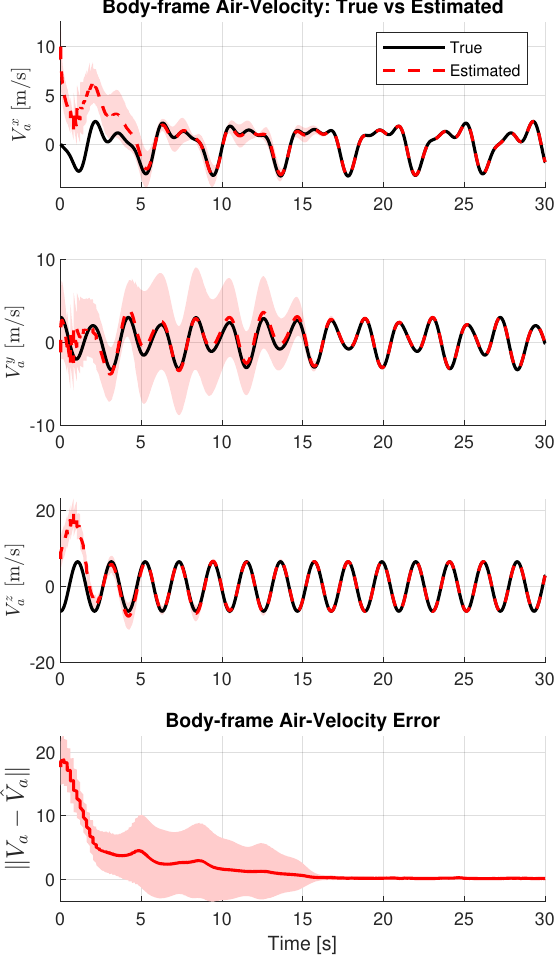}
    \caption{Estimated and True body-frame air-velocity components, and observer estimation error \(\|V_a - \hat V_a \|\).}
    \label{fig:AirVelocity_Components}
\end{figure}

\begin{figure}[!t]
    \centering
    \includegraphics[width=0.8\linewidth]{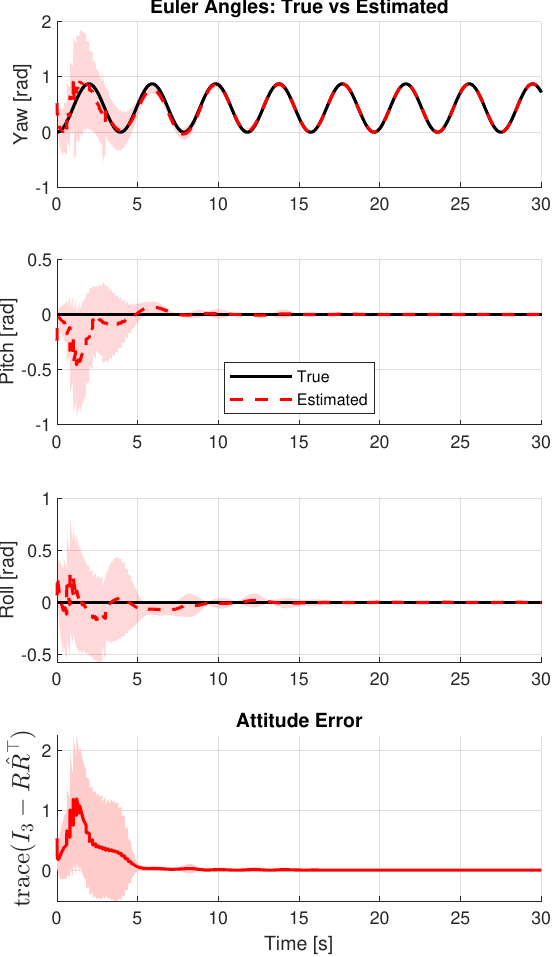}
    \caption{Estimated and True Euler angles and attitude estimation error \(\operatorname{trace}(I_3-R\hat R^\top)\).}
    \label{fig:Euler_Angles_Att_Error}
\end{figure}

A Monte Carlo simulation with 20 runs is performed, where the initial estimates are randomly sampled from Gaussian distributions. The initial body-frame air-velocity and the altitude estimates are distributed around \(\hat V_a (0) = [10,-2,8]^\top (\mathrm{m/s})\) per axis and \(\hat h (0) = 10\, (\mathrm{m})\) with a standard deviation of \(2\,(\mathrm{m/s})\) and \(1\,(\mathrm{m})\), respectively. The initial orientation estimates are normally distributed around $\hat R(0)$ which corresponds to the initial angles with a pitch of \(-\pi/20 (\mathrm{rad})\), a roll of \(\pi/20 (\mathrm{rad})\), and a yaw of \(\pi/6 (\mathrm{rad})\); with a standard deviation of $\pi/12$ per axis. The parameters are set as follows: $Q = 100*\mathrm{blkdiag}(\sigma^2_{p,1},(\sigma_m^{\top})^2,\sigma_b^2)$, $S = \mathrm{blkdiag}(0.01I_3,0.1I_3,0.01)$, $P(0)=\mathrm{blkdiag}(0.1I_3,0.25I_3,1).$ 
The accelerometer and gyroscope measurements $\mathbf{a}(t)$ and $\boldsymbol{\omega}(t)$ sampled at $200~(\mathrm{Hz})$, are corrupted with zero-mean Gaussian noise with standard deviation of $0.05$ each. The magnetometer and Pitot measurements in body frame, sampled at $50~(\mathrm{Hz})$ are corrupted with zero-mean Gaussian noise with standard deviations $\sigma_m = 0.01*[1,1,1]^\top$, and $\sigma_{p,1} = 0.5$, respectively. The barometer sampled at $5~(\mathrm{Hz})$, is corrupted with zero-mean Gaussian noise with standard deviation $\sigma_b = 0.05$. 
Figures~\ref{fig:AirVelocity_Components} and~\ref{fig:Euler_Angles_Att_Error} illustrate the air-velocity components and full air-velocity estimation error (\(\|V_a-\hat V_a\| \)), and the Euler angles and full attitude estimation error (\(\operatorname{trace}\!(I_{3} - R\hat{R}^\top)\)).
The simulation results demonstrate that the estimation errors decrease and converge to zero under Yaw rate-only motion and forward Pitot measurement. In particular, the full air-velocity and the Euler angles estimates remain bounded with fast convergence for all initial conditions. These results confirm the expected local convergence and robustness properties of the proposed observer.

\section{Conclusion}\label{sec:concl}
In this work, we proposed a nonlinear observer on $\mathrm{SO}(3)\ltimes \mathbb{R}^3\times \mathbb{R}$ for full air-velocity and attitude estimation, where the innovation terms are systematically derived from a CRE associated with the linearized error dynamics. The theoretical analysis (Lemma~\ref{Lemma1} and Theorem~\ref{Theorem1}) established a uniform observability of the augmented system and local convergence of the estimation error dynamics under persistently exciting motion. A Monte Carlo simulation with perturbed initial conditions corroborated the analysis, showing consistent convergence of the proposed observer accross all runs. Future work will extend the approach to explicitly estimate IMU-biais and wind-dynamic effects.

\appendix
\section{Proof of Lemma~\ref{Lemma1}} \label{appendix_B}
First, the corresponding observability Gramian is evaluated in a convenient block form. In particular, we block-partition the system matrix $A^{\star}(t)$ as
\begin{equation}
A^{\star}(t) =\begin{bmatrix}A^{\star}_{11}(t) & \mathbf{0}_{6\times 1} \\
A_{21} & 0
\end{bmatrix}, \label{eq:block_true_state_matrix}
\end{equation}
with
\[
A^{\star}_{11}(t) = 
\begin{bmatrix}
\mathbf{0}_{3 \times 3} & \mathbf{0}_{3 \times 3} \\
-(R \mathbf{a})^\times &  \mathbf{0}_{3 \times 3}
\end{bmatrix}, \quad
A_{21} = \begin{bmatrix}\mathbf{0}_{1\times 3}& \mathbf{e}_{3}^{\top} \end{bmatrix}.
\]
Similarly, we block-partition the output matrix $C^{\star}(t)$ as
\begin{equation}
C^{\star}(t) = \begin{bmatrix} C^{\star}_{11}(t) & \mathbf{0}_{(m+3) \times 1}\\
\mathbf{0}_{1 \times 6}  &1
\end{bmatrix}, \label{eq:block_true_output_matrix}
\end{equation}
with
\[C^{\star}_{11}(t) = \begin{bmatrix}B^{\top}R^{\top}(R V_a)^{\times} & B^\top R^\top \\
-(m_{\mathcal{I}})^{\times}& \mathbf{0}_{3 \times 3}\end{bmatrix}.
\]

Due to the structure of \(A^{\star}\) in~\eqref{eq:block_true_state_matrix}, the associated state transition matrix has the block form
\begin{equation}
\Phi^{\star}(t,\tau) =
\begin{bmatrix} 
\Phi^{\star}_{11}(t,\tau) &\mathbf{0}_{6\times 1}\\
\Phi^{\star}_{21}(t,\tau) & 1
\end{bmatrix},\label{eq:true_state_trans_matrix}
\end{equation}
with \(\Phi^{\star}_{11}\in\mathbb{R}^{6\times 6}\) and \( \Phi^{\star}_{21}\in\mathbb{R}^{1\times6}\).
From~\eqref{eq:transition_matrix} and~\eqref{eq:true_state_trans_matrix}, the derivative of the transition matrix is given by
\begin{equation}
\begin{aligned}
\frac{d}{dt} \Phi^{\star}(t, \tau)
&=\begin{bmatrix}
A^{\star}_{11}\Phi^{\star}_{11} & \mathbf{0}_{6\times 1} \\
A_{21}\Phi^{\star}_{11} & 0\\
\end{bmatrix},\quad \Phi^{\star}(\tau,\tau) = I_7. \label{eq:deriv_true_trans_matrix}
\end{aligned}
\end{equation}
with \(\Phi^{\star}_{11}(\tau,\tau) = I_6\) and \(\Phi^{\star}_{21}(\tau,\tau) = \mathbf{0}_{1 \times 6}\).
From~\eqref{eq:deriv_true_trans_matrix}, we have 
\begin{subequations}
\begin{align}
\dot \Phi^{\star}_{11} &= A^{\star}_{11}\Phi^{\star}_{11} \label{eq:dot_phi11},\\
\dot \Phi^{\star}_{21} &= A_{21}\Phi^{\star}_{11} \label{eq:dot_phi21}.
\end{align}
\end{subequations}
Due to the structure of \(A^{\star}_{11}\), \(\Phi^{\star}_{11}\) is obtained from~\eqref{eq:dot_phi11} as follows   
\begin{equation}
\Phi^{\star}_{11}(t,\tau) = \begin{bmatrix} I_3 & \mathbf{0}_{3 \times 3} \\ -\int_{\tau}^{t}(R \mathbf{a})^\times(s)ds & I_3 \end{bmatrix}\label{eq:phi_11}.
\end{equation}   
Substituting \(A_{21}\) and~\eqref{eq:phi_11} in~\eqref{eq:dot_phi21} yields:
\begin{equation}
\Phi^{\star}_{21}(t,\tau) =\mathbf{e}_3^{\top}\begin{bmatrix}
-\int_{\tau}^{t}\left(\int_{\tau}^{s}(R\mathbf{a})^{\times}\,d\sigma\right)ds &I_3(t-\tau)
\end{bmatrix}. \label{eq:Phi_21} 
\end{equation}
From~\eqref{eq:block_true_output_matrix} and \eqref{eq:true_state_trans_matrix}, we compute \(C^{\star}\Phi^{\star}\) as follows
\begin{equation}
C^{\star}(s)\Phi^{\star}(s,t)=
\begin{bmatrix}
C^{\star}_{11}(s)\Phi^{\star}_{11}(s,t) & \mathbf{0}_{(m+3)\times 1}\\
\Phi^{\star}_{21}(s,t) & 1
\end{bmatrix}\label{eq:C_Phi}
\end{equation}
Now, in view of~\eqref{eq:C_Phi} and~\eqref{eq:observability_gramian}, the Gramian is given by  
\begin{equation}
W^{\star}(t,t+\bar \delta)=
\begin{bmatrix}
W^{\star}_{11}(t,t+\bar \delta) & W_{21}^{\star \top}(t,t+\bar \delta)\\
W^{\star}_{21}(t,t+\bar \delta) & 1
\end{bmatrix} \label{eq:gramian}
\end{equation}
where 
\[
\begin{aligned}
W^{\star}_{11}(t, t+\bar \delta) &= \frac{1}{\bar \delta}\int_t^{t+\bar \delta}
\Phi_{11}^{\star \top} C_{11}^{\star \top} C_{11}^\star\Phi_{11}^\star\,ds\\ 
&+\frac{1}{\bar \delta}\int_t^{t+\bar \delta}\Phi_{21}^{\star \top}(s,t)\Phi_{21}^\star(s,t)ds,   
\end{aligned}
\]
\[
W^{\star}_{21}(t,t+\bar \delta) = \frac{1}{\bar \delta}\int_t^{t+\bar \delta}\Phi^{\star}_{21}(s,t)\,ds.
\]

To show that the system~\eqref{eq:ltv} is uniformly observable, it suffices to show that there exist \( \bar{\delta}, \bar{\mu} > 0 \) such that ~\eqref{eq:observability_gramian} holds, i.e. \(W^{\star}(t, t + \bar \delta) \ge \bar \mu I_7,\, \forall t \geq 0,\) with \(W^{\star}(t, t + \bar \delta)\) given by~\eqref{eq:gramian}.
Assume, for contradiction, that the proposed system is \emph{not} uniformly observable.  
Then, for every \(\bar \mu > 0\) and  \(\bar \delta > 0\), there exists \(t \ge 0\) such that \(W^{\star} (t,t+\bar \delta) < \bar \mu I_7\). 
Let \(\{\mu_p\}_{p\in\mathbb{N}}\) be a sequence decreasing to zero with \(\mu_p > 0\), and let 
\(\bar{\delta} > 0\) satisfy the PE conditions~\eqref{eq:PE_Pi} and~\eqref{eq:PE_a_pi}.  
Then, there exist sequences \(\{t_p\}\) and \(\{\hat d_p\}\) such that:
\(\hat d_p \in \mathcal{D} := \{\, d \in \mathbb{R}^7 : \|d\| = 1 \,\}\), the \emph{unit sphere in \(\mathbb{R}^7\)},  and  \(\hat d_p^\top W^{\star}(t_p,t_p+\bar{\delta}) \hat d_p < \mu_p, \quad \forall p \in \mathbb{N}.\)
Since \(\mathcal{D}\) is compact, there exists a subsequence of \(\{\hat d_p\}\) that converges to some \(d \in \mathcal{D}\), \(\|d\| = 1\).
Letting \(p \to \infty\) and using \(\mu_p \to 0\) yields the convergence relation from  \eqref{eq:gramian}
\begin{equation}
\lim_{p \to \infty} 
\int_0^{\bar{\delta}} 
\big\| C^{\star}\, \Phi^{\star}(s,t_p)\, \hat d_p \big\|^2 ds= 0 \label{eq:conv_relation}
\end{equation}
or, by a change of variables with an output function,
\begin{equation}
\lim_{p \to \infty} 
\int_0^{\bar{\delta}} \big\|f_p(s)\big\|^2 ds = 0, \label{eq:conv_relation_f}
\end{equation}
where we define
\[
\begin{aligned}
f_p(t) &:= C^{\star} \Phi^{\star}(t+t_p,t_p)\hat d_p\\
&= \begin{bmatrix}C^{\star}_{11} \Phi^{\star}_{11}(t+t_p,t_p) d_1\\\Phi^{\star}_{21}(t+t_p,t_p)d_1+d_2
\end{bmatrix} = \begin{bmatrix} f_{1p}(t)\\f_{2p}(t)\end{bmatrix},
\end{aligned}
\]
where \(d = [\,d_1^\top, d_2^\top\,]^\top\) with  \(d_1 \in \mathbb{R}^6\), \(d_2 \in \mathbb{R}\). This implies
\[
\begin{aligned}
\big\|f_p(t)\big\|^2 &= \big\| C^{\star}_{11}\Phi^{\star}_{11}d_1 \big\|^2 + \big\| \Phi^{\star}_{21}d_1 + d_2 \big\|^2\\
&=\big\|f_{1p}(t)\big\|^2 + \big\|f_{2p}(t)\big\|^2.
\end{aligned}
\]
Hence the convergence relation in~\eqref{eq:conv_relation_f} becomes
\begin{subequations}
\begin{align}
&\lim_{p \to \infty} 
\int_0^{\bar{\delta}} 
\big\|f_{1p}(s) \big\|^2 ds = 0, \text{ and} \label{eq:conv_PE_Pi}\\
&\lim_{p \to \infty} 
\int_0^{\bar{\delta}} 
\big\|f_{2p}(s) \big\|^2 ds = 0. \label{eq:conv_PE_dotva}
\end{align}
\end{subequations}
From~\eqref{eq:dot_phi21}-\eqref{eq:Phi_21}, we obtain the successive time derivatives of \(f_{2p}(t)\), as follow :
\[
f_{2p}^{(1)}(t) = \mathbf{e}_3^{\top}\begin{bmatrix}
-\int_{\tau}^{t}( R\mathbf{a})^{\times}ds &I_3\end{bmatrix}d_1\,,
f_{2p}^{(2)}(t) =-\mathbf{e}^{\top}_3(R\mathbf{a})^{\times}d_1
\]

Using the results of Lemma~A.1 of \cite{Pascal2017},  we deduce:
\[
\lim_{p \to \infty} 
\int_0^{\bar{\delta}} |f_{2p}^{(k)}(s)|^2 ds = 0, 
\quad k=0,1,2..
\]
Letting \(d_1 = \begin{bmatrix}d^{\top}_{1,1},d^{\top}_{1,2}\end{bmatrix}\), with \(d_{1,1} \in \mathbb{R}^3\) and \(d_{1,2} \in \mathbb{R}^3\), the highest derivative yields 
\[f_{2p}^{(2)}(t_p) \to -\mathbf{e}^{\top}_3\left( R(t_p) \mathbf{a}(t_p)\right)^{\times}d_{1,1} \to 0 \text{ as }  p \to \infty.\] This implies in the tangent-space \[-\mathbf{e}^{\top}_3\left( R(t_p) \mathbf{a}(t_p)\right)^{\times}J\bar d_{1,1} \to 0 \text{ as }  p \to \infty,\] with $d_{1,1} = \begin{bmatrix}\bar{d}{^\top}_{1,1}, \alpha \end{bmatrix}^\top$, where $\bar d_{1,1} \in \mathbb{R}^2$ and $\alpha \in \mathbb{R}$. By the PE assumption in~\eqref{eq:PE_a_pi}, this leads to \(\bar d_{1,1} = 0\).
Substituting \(\bar d_{1,1} = 0\) into \(f^{(1)}_{2p}(t_p)\), we get \(\mathbf{e}^{\top}_3d_{1,2} =0\), which implies \(d_{1,2} \in\text{span} \{\mathbf{e}_1,\mathbf{e}_2\} \), i.e., \(d_{1,2} = \begin{bmatrix}\bar d_{1,2}^\top & 0 \end{bmatrix}^\top\), where \(\bar d_{1,2} \in \mathbb{R}^{2}\).
Now, substituting $d_1 = [\alpha\mathbf{e}_3^\top, d_{1,2}^\top]^\top$, $C^{\star}_{11}$, and $\Phi^{\star}_{11}$ into $f_{1p}(t)$, we get
\(
f_{1p}(t) = \begin{bmatrix}\nu(t)\\-\alpha\,(m_{\mathcal{I}}\times\mathbf{e}_3)
\end{bmatrix},
\)
where
\[
\begin{aligned}
\nu(t) &= B^{\top}R^{\top}(RV_a)^{\times}\alpha\,\mathbf{e}_3 \\
&+ B^\top R^\top(-\alpha \int_{\tau}^{t}(R \mathbf{a})^\times(s)ds\,\mathbf{e}_3 + d_{1,2}).
\end{aligned}
\]
Hence $f_{1p}(t_p) \to 0 \text{ as }  p \to \infty$, implies $\alpha(m_\mathcal{I}\times \mathbf{e}_3) \to 0$ and $\nu(t_p) \to 0 \text{ as }  p \to \infty$.
By assumption, vectors $m_\mathcal{I}$ and $\mathbf{e}_3$ are non-collinear, then $\alpha \to 0 \text{ as }  p \to \infty$. Thus, $\nu(t_p) \to BR^\top d_{1,2} \to 0 \text{ as }  p \to \infty$. Since $d_{1,2} \in\text{span} \{\mathbf{e}_1,\mathbf{e}_2\}$ and by the PE assumption in~\eqref{eq:PE_Pi}, this implies $d_{1,2} =0$.
Substituting $d_1 = 0$ into $f_{2p}(t_p)$, we get $d_2 \to 0 \text{ as }  p \to \infty$. Hence $d = 0$, contradicting $\|d\| = 1$. Therefore, the pair $(A^{\star}(t),C^{\star}(t))$ is uniformly observable. 
\bibliography{root}  
                                    
\end{document}